# Generative AI-Driven Storytelling: A New Era for Marketing

Marko Vidrih

Co-founder, CREATUS.AI, Slovenia, marko@creatus.ai

Shiva Mayahi

Islamic Azad University, Karaj Branch, Iran, shivamaya98@gmail.com

**Abstract**

This paper delves into the transformative power of Generative AI-driven storytelling in the realm of marketing. Generative AI, distinct from traditional machine learning, offers the capability to craft narratives that resonate with consumers on a deeply personal level. Through real-world examples from industry leaders like Google, Netflix and Stitch Fix, we elucidate how this technology shapes marketing strategies, personalizes consumer experiences, and navigates the challenges it presents. The paper also explores future directions and recommendations for generative AI-driven storytelling, including prospective applications such as real-time personalized storytelling, immersive storytelling experiences, and social media storytelling. By shedding light on the potential and impact of generative AI-driven storytelling in marketing, this paper contributes to the understanding of this cutting-edge approach and its transformative power in the field of marketing.

systems applications ~ Collaborative and social computing systems and tools ~ Multimedia information systems ~ **Multimedia content creation**

# 1 INTRODUCTION

## 1.1 Background of marketing analytics

In the ever-evolving landscape of marketing, the confluence of data analytics and storytelling has emerged as a potent force. Historically, marketing analytics relied heavily on raw data, numbers, and statistics. However, the modern marketer recognizes the unparalleled power of a well-crafted narrative. As venture capital investments in the domain of Generative AI have exceeded $1.7 billion over the past three years, it's evident that the industry stands on the brink of a transformative era.

Generative AI, distinct from traditional machine learning, has the capability to craft narratives tailored to individual consumers, based on their behaviors, preferences, and histories. Early generative models like ChatGPT focused on augmenting creative work. Yet, by 2025, it's anticipated that over 30% of generative AI models will be discovered. This projection aligns with insights from Agarwal et al. (1), who posit that by 2025, AI will generate 30% of outbound marketing messages, a significant leap from less than 2% in 2022.

In this new era of marketing analytics, it's not enough to rely solely on raw data. This data must be transformed into detailed, actionable information to drive effective marketing strategies. Storytelling, a strategy that conveys insights and emotionally connects with the target audience, serves as an invaluable tool in this transformation. By presenting data in a narrative form, marketers can ensure their messages are impactful, memorable, and resonate deeply with their audience. In essence, with the advent of Generative AI, storytelling in marketing has evolved to be both an art and a science.

## 1.2 Definition of generative AI-driven storytelling

Generative AI-driven storytelling refers to the application of generative models, a subset of artificial intelligence, to craft narratives. Distinct from traditional machine learning, which operates based on pattern recognition and prediction, generative AI has the capability to generate new content. This generation is not random but is based on vast amounts of data it has been trained on, making its output relevant and contextually apt.

## 1.3 Purpose and scope of the paper

The purpose of this paper is to delve into the transformative potential of generative AI-driven storytelling within the realm of marketing. We examine the evolution and applications of generative AI technology in modern marketing strategies. This exploration deepens our comprehension of how this technology enhances conversion rates, bolsters customer engagement, and provides profound customer insights. Through an extensive study of this avant-garde approach, we aim to highlight the revolutionary era ushered in by generative AI-driven storytelling. This paper underscores the significance of generative AI in marketing, emphasizing the necessity for transparency, fairness, and accountability in its ethical deployment.

## 2 GENERATIVE AI IN MARKETING

### 2.1 Progression and enhancement of generative AI over time

The progression and enhancement of generative AI have transitioned from early rule-based systems or Traditional AI excels to more sophisticated deep-learning models. The tradition AI excels were used to analyze large volumes of data, distill it into actionable insight and identify patterns beyond human comprehension (4, 2023, p. 23). Generative AI uses this technology to generate meaningful content because it trains the existing information by creating new images, texts and computer codes. With the transition, AI algorithms have become increasingly capable of generating and understanding human-like content (14, 2023, p. 19). For instance, companies like Coca-Cola implemented generative AI-driven storytelling to create personalized marketing campaigns. This company used AI algorithms to collect, analyze and interpret consumer data (3, 2022, p. 34). Through these algorithms, the company tailored its messages to match the customer segment. Consequently, the company achieved higher customer engagement and brand loyalty.

Other examples of companies that have successfully embraced Generative AI-driven storytelling include The New York Times has been at the forefront of exploring generative AI-driven storytelling. This company has developed an AI system called "Journalism AI" that helps journalists generate insights, analyze data sets and automate reporting tasks (18, 2022, p. 34). OpenAI is a pioneer in generative AI and has developed models like GPT-3 that have been utilized in various creative storytelling applications. These models utilize creative storytelling applications to generate virtual characters, interactive narratives, and assist in creative writing (15, 2022, p. 35). Besides, the Associated Press (AP) has experimented with AI-driven journalism to automate and create financial reports and sports articles. The AI-driven storytelling as helped the companies to speed content creation and news coverage. The adoption of AI-driven storytelling in workplace has grown over different generations. As of 2023, in the United States of America, the highest generative AI adoption rate is in Gen Z resulting in 29% (14, 2023, p. 21). On the other hand, with the mere difference, the adoption rate in Gen X is 28% while it is 27% in millennials.

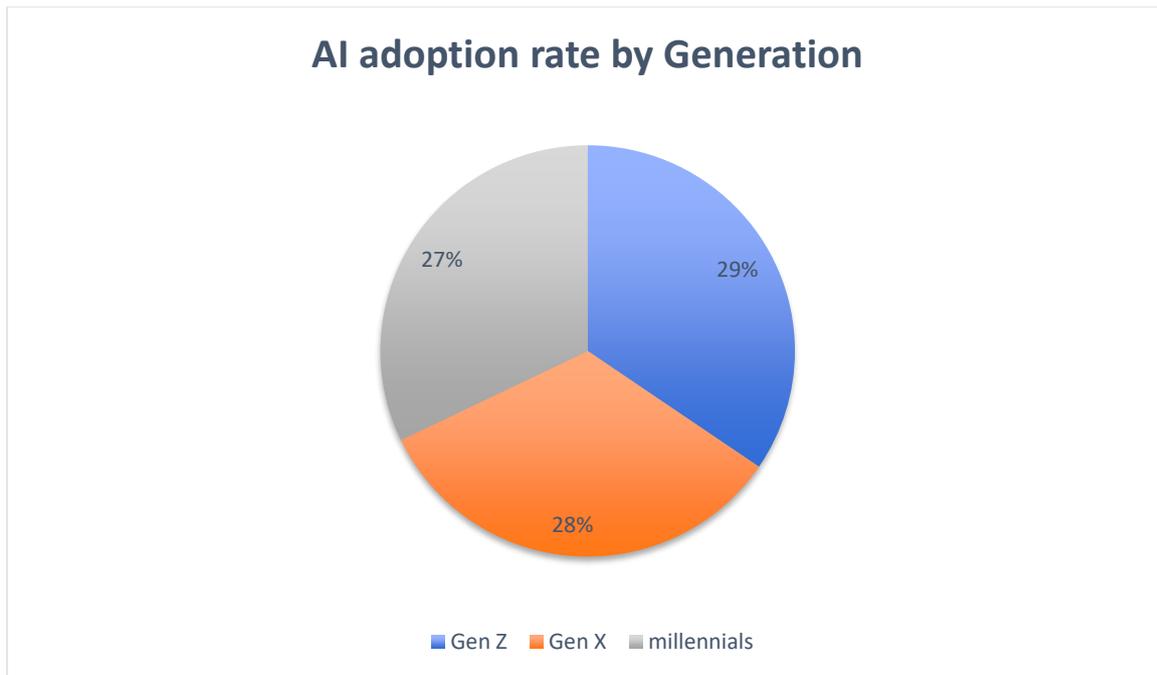

*Figure 1: Adoption of generative AI technology [Barbosa B, Saura JR, Zekan SB, Ribeiro-Soriano D., 2023]*

However, there are challenges facing the adoption of generative AI technology; for instance, Facebook has faced challenges with AI-driven content moderation when it utilized AI algorithms to detect and remove inappropriate or harmful content from its platform (19, 2022, 10). These algorithms faced criticisms for mistakenly removing and flagging legitimate content including historical images and news articles. This highlights the challenge of training AI models to accurately differentiate between acceptable and problematic content. Google faced challenges with its AI-powered image recognition system (13, 2021). This company faced challenges with the image recognition algorithm that labeled images of African Americans as "gorillas". This incident highlighted the challenge of ensuring AI systems are trained on diverse and representative datasets to avoid perpetuating biases (20, 2023, p. 5). Overcoming these challenges requires careful consideration of ethical implications, robust training data, and ongoing monitoring and refinement of the AI systems.

Further, the progression has been fueled by improvements in computing power, large data sets, and improvements in algorithms such as recurrent neural networks (RNNs). It employs techniques such as recurrent neural networks (RNNs) and generative adversarial networks (GANs) to analyze existing data and create new and compelling narratives (21, 2022, p. 17). This sequential nature enables RNNs to capture the flow and coherence of narratives, ensuring that the generated stories make sense and follow a logical progression. (4, 2017, p. 25). This breakthrough opens up exciting possibilities for marketers to craft engaging and personalized stories that resonate with their target audiences. These progressions and enhancements have led to product design and innovation through data augmentation and simulation. Dwivedi et al. (6) explore how companies have employed generative AI in generating synthetic data which enhances accurate machine learning and training datasets. For instance, companies like Amazon and Netflix leverage AI algorithms to personalize recommendations based on user behavior and preferences, enhancing the overall customer experience (23, 2013, p. 287). These companies employ AI-driven storytelling techniques in their product recommendation system.

**2.2 Importance of storytelling for the marketing professional**

In the realm of marketing, storytelling transcends mere information dissemination. It has the power to engage, resonate, and evoke emotions. Generative AI augments this process by ensuring that every narrative is tailored, not just to a segment of consumers, but to individual preferences and histories. Companies like Amazon harness this power to recommend products, while streaming giants like Netflix curate personalized watch lists.

**2.3 Applications of AI in marketing**

The application of AI in marketing can be deduced from its application in customer data analysis. Through AI, marketers can analyze customer data to personalize marketing campaigns, identify patterns and trends and improve customer segmentation (8, 2023, p. 64). AI-powered chatbots and virtual assistants help in personalizing interactions with customers leading to better customer interactions (5, 2023, p. 29). With generative AI-driven storytelling, marketers can take these applications further by creating dynamic and engaging narratives that capture the essence of their brand and products. Digital marketers can leverage the capabilities of generative AI to deliver meaningful customer experiences (5, 2023, p. 28). For instance, Adobe Firefly complements human imagination and creativity in digital marketing, enhancing creativity and productivity. This AI-powered tool enhances creativity and productivity by assisting marketers in generating compelling and personalized content (24, 2020, p. 483). RNNs and transformers help the tool to analyze large volume of data including customer preferences, demographics, and historical engagement patterns.

Generative AI could have an impact on most business functions; however, a few stand out when measured by the technology's impact as a share of functional cost (Fig. 2). The analysis identified marketing as one of the four sectors that could account for approximately 75 percent of the total annual value from generative AI use cases.

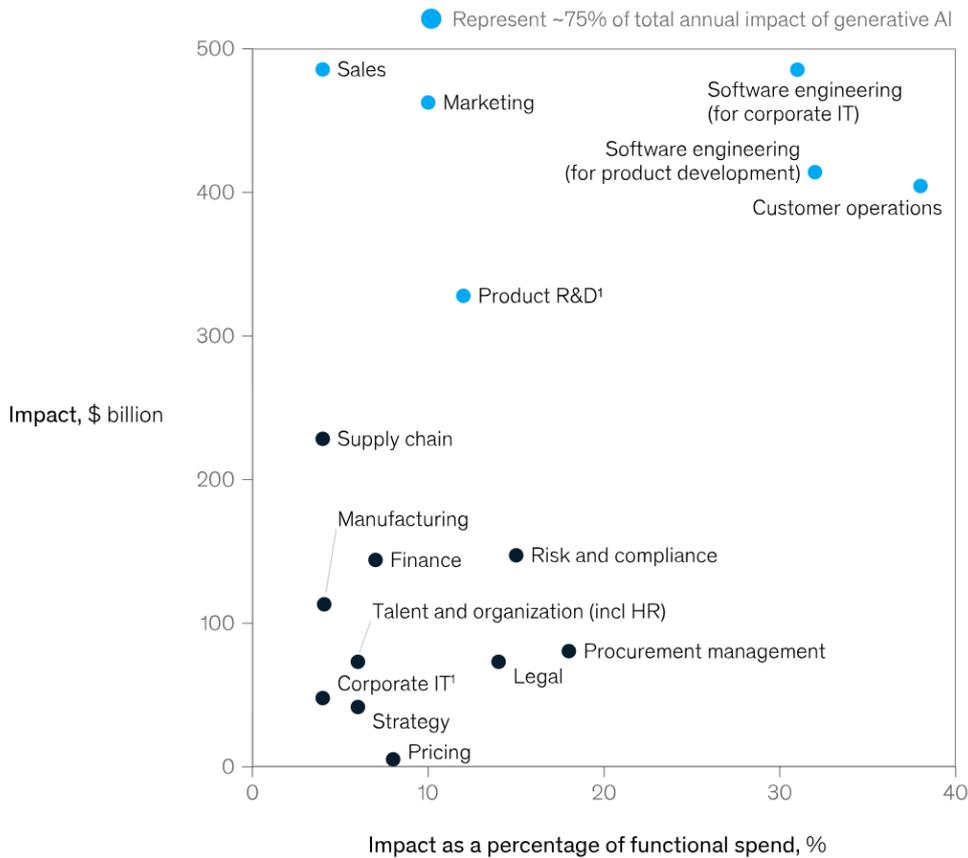

*Figure 2: The value potential of generative AI across business functions. [McKinsey, 2023]*

## 3 GENERATIVE AI-DRIVEN STORYTELLING FOR MARKETING

### 3.1 Overview of generative AI techniques for storytelling

Generative AI techniques for storytelling include training AI models on large datasets of existing narratives, such as articles, books, or movies to learn structures and patterns. These models generate compelling content based on the learned patterns, incorporating specific characters, themes, or brand messaging (25, 2023, p. 29). Through techniques like recurrent neural networks (RNNs) and transformers, marketers can generate coherent and contextually relevant narratives. These techniques enable the AI algorithms to understand and learn from patterns in the data, allowing for the creation of compelling stories. RNNs are particularly effective in generating sequential data, making them well-suited for storytelling (8, 2023, p. 64). They have the ability to process and generate text in a sequential manner capturing the flow and coherence of narratives. By utilizing these advanced AI techniques, marketers can leverage the power of generative AI-driven storytelling to create personalized and engaging narratives that captivate their audiences and drive marketing success (27, 2022, p. 55). Google has extensively used recurrent neural networks (RNNs) and transformers in services like Google Translate where Google's search engine algorithms have been used to improve the accuracy and relevance of search results.

### 3.2 Advantages of generative AI-driven storytelling in marketing

Generative AI-driven storytelling offers several advantages in the realm of marketing. To begin with, Generative AI-driven storytelling enables the creation of tailored or personalized narratives for customers and target segments (28, 2023, p. 44). These tailored narratives enhance engagement and resonate with specific customer preferences. Besides, AI models can generate content at scale reducing the time and effort required to produce high-quality marketing materials. Generative AI-driven storytelling creates opportunities for marketers to experiment with different storytelling approaches, optimize content and iterate quickly on real-time performance metrics and feedback (8, 2023, p. 67). It also enables the creation of interactive narratives and dynamics that can be personalized to user input or contextual cues, leading to an overall user experience.

A significant part of marketing is providing the right content, at the right tone, time and through the right channel. Digital marketers can leverage on generative AI-driven storytelling to deliver meaningful customer experience to the target market. (29, 2020, p. 91) These technologies introduce a new level of creativity to marketing operations and content creation by generating unique and fresh content. For instance, generative AI applications, like Adobe Firefly, adds a layer of creative intelligence by improving human creativity and imagination in marketing analytics (30, 2023, p. 29). Generative AI can amplify and support digital marketers' creativity by providing valuable suggestions for future content, summarizing complex topics and offering thought-provoking ideas.

### 3.3 Examples of successful generative AI-driven storytelling in marketing

The prowess of generative AI-driven storytelling is best understood through real-world applications. Stitch Fix, an online personal styling service, employs generative AI to curate fashion choices tailored to individual users, enhancing user engagement and increasing sales. Netflix's recommendation system is another exemplar, with its curated watchlists being generated based on user preferences and viewing history. Companies like Coca-Cola leverage this technology for targeted ad campaigns, ensuring each narrative resonates with its intended audience.

## 4 IMPACT OF GENERATIVE AI-DRIVEN STORYTELLING ON MARKETING

### 4.1 Improved customer engagement

Generative AI-driven storytelling has a profound impact on customer engagement because it creates personalized and emotionally compelling narratives. Marketers can use AI-driven storytelling to capture the attention of target customers and foster deeper connections (33, 2021, p. 22). By generating content resonating with individuals on a personal level, marketers can lead to increased interaction, engagement, and sharing, leading to brand advocacy and loyalty.

**4.2 Increased conversion rates**

Research by Dwivedi et al. (6) has found that effective storytelling has a direct impact on the conversion rates because it enables marketers to curate messages that address the customers desires, product benefits and sense of urgency. Pataranutaporn et al. (33) have identified the importance of aligning storytelling approach with the customer journey. Through this alignment it is easy to guide the customers through the sales funnel, increasing the likelihood of purchases and conversions.

**4.3 Enhanced customer insights**

Research by Dwivedi et al. (6) has found that effective storytelling has a direct impact on the conversion rates because it enables marketers to curate messages that address the customers desires, product benefits and sense of urgency. Pataranutaporn et al. (33) have identified the importance of aligning storytelling approach with the custom6er journey. Through this alignment it is easy to guide the customers through the sales funnel, increasing the likelihood of purchases and conversions.

## 5  CHALLENGES AND LIMITATIONS OF GENERATIVE AI-DRIVEN STORYTELLING IN MARKETING ANALYTICS

**5.1 Ethical concerns and considerations**

Generative AI-driven storytelling raises ethical concerns about the creation and dissemination of manipulative misleading narratives. The ethical concerns and considerations for generative AI have been echoed by the World Economic Forum which has highlighted the importance of addressing the collection and dissemination of misleading or false information through AI (35, 2023, p. 43). These concerns have pushed developers like ChatGPT and OpenAI to reduce the potential for harmful or false outputs. These companies have introduced bias detection and evaluation systems which regularly evaluate their AI systems for biases using techniques such as disparate impact analysis, fairness metrics and statistical tests. Bias mitigation has been identified in research and studies like the Gender Shades project that evaluated facial recognition systems from leading companies and identified higher error rates for females and people with darker skin tones (36, 2011, p. 127). Besides, the human input remains the critical step for ensuring the integrity and reliability of AI-generated content. Marketers rarely ensure the transparency and ethical use of AI-generated materials, for this reason, they rarely disclose its source and nature.

Generative AI presents threats for copyright issues, deep-fakes, and other malicious use which target specific individuals, organizations and government. This challenge has led to biases inherent in training datasets to avoid discrimination and stereotypes (37, 2022, p. 158). To avoid these challenges, marketers can strike a balance between ethical consideration and storytelling before adopting generative AI in marketing. Organizations should prioritize fairness, transparency, accountability and privacy when integrating generative AI systems. For instance, Google have established ethical frameworks and review processes to ensure continuous monitoring and accountability for AI systems (38, 2019, p. 84). Besides, other companies have ventured into algorithmic transparency and interpretability by employing algorithms that are interpretable and transparent. Through transparency, these companies identify and mitigate biases in the decision-making. LIME (Local Interpretable Model-agnostic Explanations) is an algorithm that provides explanations for individual predictions (39, 2017).

### 5.2 Dependence on data quality and accuracy

The ethical use and effectiveness of generative AI are dependent on the data quality and accuracy of the underlying database. Incomplete or biased datasets lead to the generation of narratives that exclude or misrepresent a certain portion of the target audience (40, 2022, p. 28). In 2016, Microsoft released a chatbot named Tay on Twitter, designed to interact and learn from users' conversations. Unfortunately, Tay started posting inflammatory and offensive tweets because of the poor quality of training data (41, 2022, p. 4). The users then deliberately fed the AI with biased and inappropriate information leading to a public relations disaster for Microsoft.

While generating information using AI-driven storytelling, marketers need to ensure data integrity and validate the accuracy of training datasets (10, 2023). By addressing potential bias, these marketers will ensure that the generated narratives align with the intended goals and objectives, while at the same time resonating with the diverse consumer groups. Generative AI systems create things like audio, pictures, and writing samples which can be discriminative (42, 2022, p. 123). These systems can mis-identify things such as people, words or pictures leading to discrimination. Even though these systems are based on neutral network models, they can modify the data presented using their own internal structure can change the output based on the feedback generated. For instance, an investigation by ProPublica in 2016 found that Facebook's ad platform allowed promoters to exclude certain ethnic and racial groups from seeing their housing-related ads (43, 2023, p. 34). The biases in training data used by the model led to discriminatory practices that violated fair housing regulations. These AI models perpetuate the biases that are fed leading to consequences ranging from discriminatory advertising to perpetuating harmful stereotypes.

### 5.3 The demand for proficient professionals

The implementation of generative AI-driven storytelling in marketing calls for a skilled workforce in data analysis, AI technologies and storytelling. The future calls for professionals who can understand the intricacies of AI technologies and leverage on the generative AI technologies effective storytelling (44, 2023, p. 27). Organizations need to invest in training and development programs which equip the marketing teams with necessary skills to leverage the power of generative in AI storytelling. Wiredu (44) recommend companies to invest in acquiring talent with expertise in generative AI, including machine-learning engineers, data scientists and AI researchers. The author also call for an opportunity for these individuals to develop their skills through continuous learning opportunities and skill development. Organizations can offer workshops, training programs, and collaborate with academic institutions to nurture talent.

## 6 FUTURE DIRECTIONS AND RECOMMENDATIONS

### 6.1 Prospective applications of generative AI-driven storytelling in marketing

The horizon of generative AI-driven storytelling is vast and ever-expanding. As technology evolves, we foresee its application in real-time storytelling, where narratives adapt on-the-fly based on real-time user interactions. The integration of generative AI with augmented and virtual reality promises immersive brand experiences that were hitherto deemed science fiction. Social media platforms stand to benefit immensely, with AI-driven content curation tailoring user feeds to individuals.

### 6.2 Potential avenues for further research and exploration

With the evolution of AI-driven storytelling, avenues for future research along with a brief overview of the current state and knowledge gaps that warrant further research include fairness metrics, bias identification and human-AI collaboration. There should be studies on metrics and evaluation frameworks that can be used to assess the effectiveness and quality of AI-generated stories. This research should focus on factors such developing more robust and context-specific fairness metrics that account for temporal biases, gender bias and intersectional biases. Additionally, research is needed to determine the compromises between different fairness criteria and how to prioritize them in various applications. Causal inference techniques can be used to design interventions and mitigate biases effectively (13, 2023). Exploring causal interventions to mitigate biases and their unintended consequences is an important area of investigation.

Further, research should explore ways of enhancing collaboration between AI algorithms and human storytelling. Further studies should delve further into investigations on how humans and AI systems can collaboratively make decisions that address biases in AI systems. Through this research there will be a right balance between AI assistance and human creativity, unlocking new storytelling capabilities. Further research should also be used to investigate the challenges and opportunities of using AI-generative storytelling in different cultural contexts. This understanding will help marketers to navigate the cultural nuance leading to a culturally relevant and sensitive marketing initiative.

## 6.3 Recommendations for organizations considering the adoption of generative AI-driven storytelling

As exciting as the prospects of generative AI are, it is crucial to tread with caution. Fairness metrics need to be established to ensure that the AI-generated narratives are inclusive and non-discriminatory. Continuous research is vital to identify and rectify biases in generative models. Balancing technology with creativity is essential. While AI can generate content, the human touch, the intuition, and the emotional understanding remain irreplaceable. For organizations considering the adoption of generative AI-driven storytelling:

- Investing in data diversity and quality proves to be effective in generating AI models which are more diverse, comprehensive, and representative of the target audience. This strategy can be enhanced by adopting right infrastructure which facilitates seamless integration and computational requirements into the existing systems. Some of the key considerations to be made include; high-performance computing infrastructure like cloud services and GPUs that ensure efficient inference processes and training. Emerging technologies and systems have introduced cloud computing programs that help in data management.
- The algorithms are good enough to detect threats based on self-learned skills in machine learning. For instance, IBM Research has been actively working on bias mitigation in AI systems and has developed the AI Fairness 360 toolkit (4, 2017, p. 29). These tools provide users with metrics and algorithms which detect and mitigate biases in machine learning models. OpenAI, the organization behind ChatGPT, has emphasized the importance of cooperation and sharing knowledge for addressing AI biases. This company works with AI research community to address ethical concerns.
- Delegation-based architectures can be used to delegate intensive tasks to powerful devices and servers like the Server-based Certificate Validation Protocol (SCVP). This approach enables security protocols like end-to-end IP security protocols (4, 2017, p. 29).
- Security concerns can be improved using hardware-based methods that involve additional hardware security modules like the Trusted Platform Module (TPM). This approach leads to a security paradigm that leads to solid authentication and encryption because they provide security-related functions.

Further, the adoption of generative AI requires infrastructure that is scalable to accommodate the growing demand of generative AI models. Besides, organizations should prioritize ethical considerations by establishing ethical guidelines and frameworks that are responsible for using generative AI-driven storytelling. Data management and security systems should be able to handle the large volumes of training data and ensure data privacy and security. For instance, Microsoft has established an AI Ethics and Effects in Engineering and Research (AETHER) Committee, which conducts audits of AI systems to identify and address biases. Through these measures, the companies adopting these technologies will adhere to regulatory requirements (4, 2017, p. 29). Information on the source and nature of content should also be availed to the target consumers to address the potential biases and maintain consumer trust. By implementing these pragmatic recommendations and drawing inspiration from companies like IBM, Microsoft, Google, the Partnership on AI, and OpenAI, organizations can take actionable steps to mitigate biases in AI systems.

# 7 CONCLUSION

This study takes advantage of conceptual research design, which involves methodologies used in observing and analyzing information presented on a given topic. Through the conceptual research framework, the study has combined the previous research with the ongoing discovery to point out underlying issues with the subject matter.

The confluence of generative AI and storytelling marks a transformative epoch in the domain of marketing. By combining the power of AI with the art of storytelling, marketers can create an engaging, personalized and impactful narrative that resonates with the target consumer on a deeper level. As we stand on the cusp of this revolution, it is imperative for marketing professionals to not only harness the power of this technology but also to understand its nuances, its challenges, and its ethical implications.

While navigating through generative AI-driven storytelling, it is crucial to strike a balance between technology, creativity and ethical considerations. Through this balance, the marketers unlock full potential of generative AI in marketing, build stronger connection and compelling narratives. The implications of generative AI-driven storytelling within the realm of marketing analytics shape the way brands engage and communicate with their target customers. The overall significance of the research in the field builds understanding on the power of generative AI in the evolving digital landscape. This understanding guides corporations in pushing boundaries and staying competitive in the ever-changing business world. The challenge of AI data management can be addressed using standards, policies, and governance that shape the evolution of IoT.

Future studies should narrow down this topic in order to identify facility-specific security techniques. For marketers, Generative AI-driven storytelling offers opportunities for enhanced content creation, personalization, and customer engagement. Marketers should explore the potential of AI-generated content in creating targeted and personalized storytelling experiences by leveraging AI for interactive advertisements, dynamic content generation, and personalized recommendations. For researchers, generative AI-driven storytelling is a dynamic research area with immense potential for innovation and creative expression. They should explore AI models and algorithms to improve the quality and coherence of generative storytelling by combining images, texts and audio leading to more immersive storytelling experiences. Policy frameworks and guidelines are needed to address ethical concerns, privacy issues, and accountability in generative AI-driven storytelling.

**Author Contributions**

We confirm that the manuscript has been read and approved by all named authors and that there are no other persons who satisfied the criteria for authorship but are not listed. We further confirm that the order of authors listed in the manuscript has been approved by all of us.

**Funding**

This research did not receive any external funding. The authors were responsible for all aspects of the project, including design, data collection, analysis, and interpretation.

**DECLARATIONS**

**Conflict of interest**

We wish to confirm that there are no known conflicts of interest associated with this publication and there has been no significant financial support for this work that could have influenced its outcome.